\documentclass[aps,floatfix,amssymb,preprint]{revtex4}

\usepackage{amssymb}
\usepackage{amsmath}
\usepackage{epsfig}
\usepackage{graphics}
\newcommand{\bn}[1]{\mbox{\boldmath $#1$}}
\usepackage{graphicx}
\usepackage{amsxtra}

\begin{document}

\title{Tuning Fano-type resonances in coupled quantum point contacts by applying asymmetric voltages  \vspace{-0.8cm} }

\author{Rub\'en C. Villarreal$^{1,2}$}
\author{Francisco  Mireles$^{2}$}
\author{Ernesto E. Marinero$^{3}$}
\author{Bruce A. Gurney$^{3}$}
\affiliation{$^{1}$Centro de Investigaci\'on Cient\'ifica y de Educaci\'on Superior de Ensenada, Ensenada, BC, 22800, M\'exico}
\affiliation{$^{2}$Centro de Nanociencias y Nanotecnolog{\'{i}}a, Universidad Nacional Aut\'onoma de M\'exico, Ensenada, BC, 22800, M\'exico}
\affiliation{$^{3}$Hitachi San Jose Research Center, Hitachi Global Storage Technologies, San Jose CA 95135, USA}
\date{\today}

\begin{abstract}
We study the ballistic magnetotransport in a double quantum point contact (QPC) device consisting of a quasi-one-dimensional quantum wire with an embedded island-like impurity - etched nano-hole as in a recently published experiment [J. C. Chen, Y. Lin, K.-T. Lin, T. Ueda and S. Komiyama, Appl. Phys. Lett. {\bf 94}, 012105 (2009)]. We reproduce the zero field quantized conductance, the interference phenomenon induced by the coupled QPCs, as well as the Ramsauer-like resonances observed in the experiments. At finite magnetic fields Fano-type resonances arises in the conductance due to the formation of localized states at the impurity periphery and to an inter-edge state resonant coupling effect. It is predicted that the Fano-type resonances can be controlled by an asymmetric confinement of the QPCs.
\end{abstract}

\maketitle
The discretization of the conductance in units of $2e^2/h$ of a quantum point contact (QPC), with $-e$ the elementary electric charge and $h$ Planck's constant, was beautifully demonstrated more than two decades ago. \cite{van Wees,Wharam} The effect was soon understood within the noninteracting electron picture. \cite{Buttiker, Beenakker,Castano-Kirczenow} Nowadays QPCs and narrow short wires are still one of the most inspiring experimental devices to study fundamental physics and new transport phenomena. Among them, the role of (charged) impurities and disorder, \cite{Ando,Bagwell, Ferry, Nonoyama1, Zozoulenko} the effect of electron-electron interactions and spin related phenomena, such as the Kondo effect, \cite{Meir, Rejec} spin-orbit interaction, \cite{Mireles-Kirczenow} and the still puzzling origin of the 0.7 anomaly, \cite{Thomas} to mention just a few. Undoubtedly, its understanding would shed light on the implementation of nanoscale sensors and spintronic devices.   For instance, exploiting the control of the symmetry of the QPCs confining potential has been recently proposed as an all-electric experimental alternative to create spin polarized currents.
\cite{Debray}

Recently Chen \textit{et al}. \cite{Chen} reported interesting measurements of the ballistic conductance of a device formed by a parallel double QPC defined in a narrow quantum wire by gating a two-dimensional electron gas (2DEG) and chemically etching a hole (of diameter $d\geq\lambda_{F}$, with $\lambda_{F}$ is Fermi wavelength) that functions as an island-like impurity scatterer.  
A full theoretical description of such experiment is absent and providing a detailed modeling framework is therefore desirable. Moreover, the understanding of the influence of a magnetic field on the conductance of such type of device is certainly of great interest for applications in nanoscale magnetic sensors and scanning probe devices.

In this letter we present a theoretical simulation of the conductance results provided by Chen {\it et al.}\cite{Chen} performed at zero field.
The main features of the conductance observed in the linear regime of such device are well explained within our recursive Green's function approach. Modeling the gate controllable electrostatic width of the QPCs allow us to study the quantization and distortion of the  conductance, as well as the Ramsauer-type and inter-channel interference effects of electrons emerging from the coupled QPCs as seen experimentally.  Our theoretical analysis validates in general the experimental results with the exception of a slight discrepancy when both QPCs are open. The possible reasons for such discrepancy are also discussed.
Additionally we predict that the presence of a magnetic field perpendicular to the 2DEG favors the formation of nearly bound states\cite{Takaferry,Jain} around the etched hole of the device. These in turn traps electrons via an interedge state coupling mechanism, strongly inhibiting transport. It is shown that such behavior is tunable by applying an asymmetric voltage gating.

The starting point of our approach is a single particle tight-binding Hamiltonian in a  two dimensional lattice model, with $
H=\sum_{\bn r}\varepsilon_{\bn r}|\bn r\rangle \langle \bn r|-\sum_{\bn r,\delta \bn r } V_{\bn r, \delta \bn r }|\bn r\rangle \langle \bn r +\delta \bn r |,
$
where $|\bn r\rangle \equiv |m\,n\rangle$, with $n$ and  $m$ are the $x$ and $y$ lattice points, respectively, $\bn r +\delta \bn r$ is the nearest neighbor site to $\bn r$, $\varepsilon_{\bn r}=U_{\bn r}+4t$ is the effective onsite energy with $U_{\bn r}\equiv U_{mn}=U(x=na,y=ma)$ the net local potential and $V_{\bn r, \delta \bn r }=\hbar^{2}/2m^{*}a^{2}=t$ is the hopping parameter, being $m^*$ the electron effective mass  and $a$ the lattice constant. The external magnetic field $B$ is taken into account by the usual Pierls substitution which introduces  a phase factor in the hopping parameter by an appropriate selection of the Landau gauge, ${\bn A}=(-yB,0,0)$, such that $V_{nm,n-1\,m^{\prime}}=-t e^{-2\pi i\Phi\left[m-(M+1)/2\right]}\delta_{mm^{\prime}}$ where $\Phi=Ba^{2}/\phi_{0}$, $\phi_{0}=h/e$ is the magnetic flux quantum and $M a$ is the width of the channel. 

The physical system comprises a quasi-one dimensional wire parallel to the $x-$axis attached to semi-infinite ideal leads with hard wall boundary conditions. The central region of the quantum wire (scattering region) contains three types of potentials. One that defines the central impurity $V_{imp}(x,y)$ (modeling the etched-hole of Ref.[\onlinecite{Chen}]) and two smooth confining potentials that in conjunction with $V_{imp}(x,y)$ and relative to the sidewall of the quantum wire, creates two independent parallel QPCs ($V_{QPC_{1,2}}$), as employed in the experiment. A suitable potential profile to describe the QPCs 
is modeled using $V_{QPC_i}(x,y)= V_{i}\cos^2\left( {\pi x}/{L_{x}}\right) +E_{F}\sum_{\pm}\left(\frac{y-y_{\pm}(x)}{\Delta}\right)^{2}\Theta(y^{2}-y_{\pm}(x)^{2})$  where $\Theta(\mu)$ is the unit step function ($\Theta=1$ for $\mu>0$ and $\Theta=0$ otherwise), and  $y_{\pm}(x)=\pm\frac{L_{y}}{2}\sin^2\left(\pi x/{L_{x}}\right)$. \cite{Ando} The parameter $\Delta$ characterizes the effective width of the QPCs, $V_{i}$ represents the maximum bottom energy at $x=y=0$ for a given QPC$_{i}$ ($i=1,2)$, $L_{x},L_{y}$ are the length and width of the wire respectively and $E_{F}$ is the Fermi energy.

The linear-response conductance is determined within the Landauer framework of ballistic transport which is appropriate if the electron-electron interactions are neglected. The conductance is expressed in terms of the transmission amplitudes by the multimode formula $G=(2e^{2}/h)\sum_{l,l^{\prime}}t_{l,l^{\prime}}t_{l,l^{\prime}}^{*}$, where the summation goes over all the propagating channels at the leads at a constant Fermi energy.\cite{Landauer} 
The transmission amplitude is calculated via the Fisher-Lee's formula, \cite{Lee-Fisher}  $t_{l,l^{\prime}}=i\hbar\sqrt{v_l /v_{l^\prime}}{\cal G}^{l,l^{\prime}}_{N+1,0}$, in which $v_{l}(v_{l^{\prime}})$ is the velocity of the outgoing (incoming) electrons. The total Green's function ${\cal G}^{l,l^{\prime}}_{N+1,0}$, connects the left and right leads and it is computed by a standard recursive technique using the Dyson's equation. \cite{Ferry-Book}


In our simulations the experimentally etched nano-hole is modeled through a circular disk scatterer of diameter $d=75$\,nm with infinitely high repulsive potential ($V_{imp} \gg E_F$).  We use a Fermi wavelength of $\lambda_{F}=50$\,nm (with $\lambda_{F}/a=4$) which is a typical value for a 2DEG formed in a GaAs/AlGaAs semiconductor heterostructures which yields  $d\geq\lambda_{F}$, consistent with values in the experiment of Chen et al. Ref.[\onlinecite{Chen}].  The effective aperture of the QPCs can be controlled independently by tuning $V_{1(2)}$ in our modeling. All energies are in units of $E_F$.

In Fig. 1({\it a}) we show the numerical results for $G$ versus $V_2$ for different fixed values of $V_1$ at QPC$_{1}$  as well as the symmetric case $V_2=V_1$ (all with field $B=0$). For $V_1=1.0$ and $V_2>0.8$ the conductance is zero indicating that both QPC$_{1(2)}$ are closed. When $V_2<0.8$ with  $V_1=1.0$ transport is allowed only through QPC$_2$ and a step-like conductance in units of $\sim 2e^2/h$ with weak resonance signals superimposed is obtained, in remarkable similarity with the measurements of Ref.[\onlinecite{Chen}] (see Fig. 2.({\it a}) in that reference). Note that the inflection point at $V_2 \sim 0.75$ and the oscillatory behavior for $V_2<0.5$ are in reasonably good agreement with the experimental results. 
The feature at $V_2\sim0.75$ occurs at $G=e^2/h$ but should not be confused with a spin-polarized transport as the leads were assumed to be spin-unpolarized and exchange-correlation effects were not considered here. The latter behavior is likely due to the abrupt potential profile at the vicinity of the central impurity causing Ramsauer-like interference phenomena. As $V_1$ is decreased propagating modes through QPC$_1$ begin to participate in the transport. In contrast to the experiment, when both QPCs are open, we observe that the resonance features get smeared out producing a rather smooth staircase profile for $G$. A possible explanation for such discrepancy may be associated with difficulties in controlling experimental parameters such as the abrupt change of the confining potentials as the QPCs are opened, or due to strong irregularities at the periphery of the central scatterer possibly produced by the chemical etching. The symmetric case ($V_2=V_1$) shows the expected classical addition of the conductances ($G\simeq 2G_2$, with $G_1\simeq G_2$), that  occurs due to a coherent superposition of the propagating modes through the parallel double QPC device exhibiting plateaus at $G=4e^2/h$ as the applied voltage ($V_2/|e|$) is decreased. 

In Fig. 1({\it d}) we depict the conductance map $G$ in the $V_1-V_2$ plane. The dark color region at $V_{1(2)}\gtrsim 0.8$ represents the condition of maximum opacity of the QPCs, while the light color region for $V_1$ and $V_2\lesssim 0.1$ characterizes the maximum conductance. Due to the symmetric geometry of the device, a mirror image of $G$  with respect to the trace at $V_1=V_2$ is obtained forming a tile-like mapping instead of the diamond-like structure seen in the experiments,\cite{Chen}  the latter due presumably to the slight coupling of the capacitance between the two gate electrodes.  To check this picture we calculated the conductance with a simple capacitance model in the linear response regime which provided good agreement with experiment (inset ({\it e}) of Fig. 1).\cite{capacitance}


Next we proceed to examine the influence of a perpendicular magnetic field on the conductance of the device.
In Fig. 2({\it a}) we show traces of $G$ vs the magnetic flux $\Phi$ (in units of $\phi_0$) for typical values of $V_1$ and $V_2$ in the regime for which up to two propagating modes are allowed to emerge from both QPCs. The  plots are vertically shifted for the sake of clarity. The behavior of the conductance shows remarkable features. First, at low fields ($\Phi\lesssim 0.05$) a set of resonances pattern arises, which are surprisingly rather insensitive to the interplay and strength of $V_1$ and $V_2$. 
 Second, for moderated magnetic flux intensities in the range $0.05\lesssim \Phi\lesssim 0.12$ a roughly periodic antiresonance (dips) of the conductance are visible. In drastic contrast to the case with low fields, in the present regime the behavior of the anti-resonances are strongly affected by $V_1$ and $V_2$, although its period is basically constant at all voltages. 
The period of the oscillations agrees well with the theory of the Aharonov-Bohm (AB) effect. According to the AB effect, the periodicity of the oscillations satisfy $S\Delta\phi=a^2\phi_{0}$,  where $S$ is the area  of the impurity (antidot), and $\Delta\phi$ is   related to the addition of one flux quantum $\phi_{0}$ to the total flux $\phi=BS$ as $B$ is increased. From the period of the antiresonance oscillations we can extract an effective surface area $S^*$ of a circular loop  of diameter $D^{*}$ of about $129\,$nm.  From the plot (Fig. 2({\it c})) of the local density of states (LDOS) calculated via $\rho(\bn r;E)=-\frac{1}{\pi}{\text{Im}}\,{\cal G}({\bn r},{\bn r};E)$,    we can define a circular contour of high electronic density around the impurity of an effective diameter of $D\sim 134\,$nm,  in quite good agreement with that inferred from the AB oscillations and with the cyclotron radius $r_c=2\pi\hbar/e B\lambda_F = 72.9\,nm$ at $\Phi=0.042$.   Fig. 2({\it b}), 2({\it c}) and 2({\it d}) shows the LDOS for the symmetric gating ($V_1=V_2 = 0.62$). The LDOS plot of Fig. 2({\it b}) corresponds to  $\Phi=0.06$ for which $G=2e^2/h$ while Fig. 2({\it c}) corresponds to magnetic flux ($\Phi=0.066$) location for which a vanishing (dip) conductance is obtained, whereas Fig.2({\it d}) is the LDOS for the resonance at $\Phi=0.042$ with $G\sim 4e^2/h$. 
The evolution of the conductance from a resonance  to an anti-resonance pattern as $\Phi$  is increased can be intuitively understood within the framework of magnetic edge state theory via an inter-edge resonant coupling mechanism as studied early in other devices. \cite{Kirczenow-etal,Takaferry}
For instance, in the  case of Fig.2({\it b}) a resonant tunneling between the left and the right higher edge states can occur via the circular edge states around the etched hole. This gives rise to the periodic resonances of $G$. On the other hand, in Fig.2({\it c}) the electrons are strongly localized around the disk periphery with a very low density of the right edge-states. This explains the periodic sudden drop of the conductance, induced by the resonant coupling (reflection) between these circular magnetic bound states with the left and right moving edges states of the wire. On the other hand, increasing the assymetry\cite{asymmetry} of the QPCs ($V_1 \ne V_2$) drastically diminishes the depth of the dips (Fig.2({\it a})).   Hence an interesting switching effect at a fixed $\Phi$ due to the asymmetric tunable electrostatic width of the QPCs may also occur.  A reasonable explanation of the physical origin of the effect can be also derived within the edge-state coupling picture (Fig.2.({\it e})) Such kind of behavior could be observed in typical GaAs/AlGaAs as based devices at low temperatures and relatively low magnetic fields strengths ($\sim 1$T).

In sumary, we have reproduced qualitatively the main experimental features of the (linear-response) conductance measurements of a double QPC device carried out at zero field. At finite magnetic fields we  predict that the formation of localized states around the  impurity induces  Fano-type resonances that strongly inhibits electron transport. Interestingly, the Fano-type resonances can be tuned by an asymmetric voltage gating of the device at low magnetic fields and cryogenic temperatures.

\begin{figure}[!]
\vspace{-3.0cm}

\hspace*{3.0cm}\includegraphics[width=5.5in,height=7.3in]{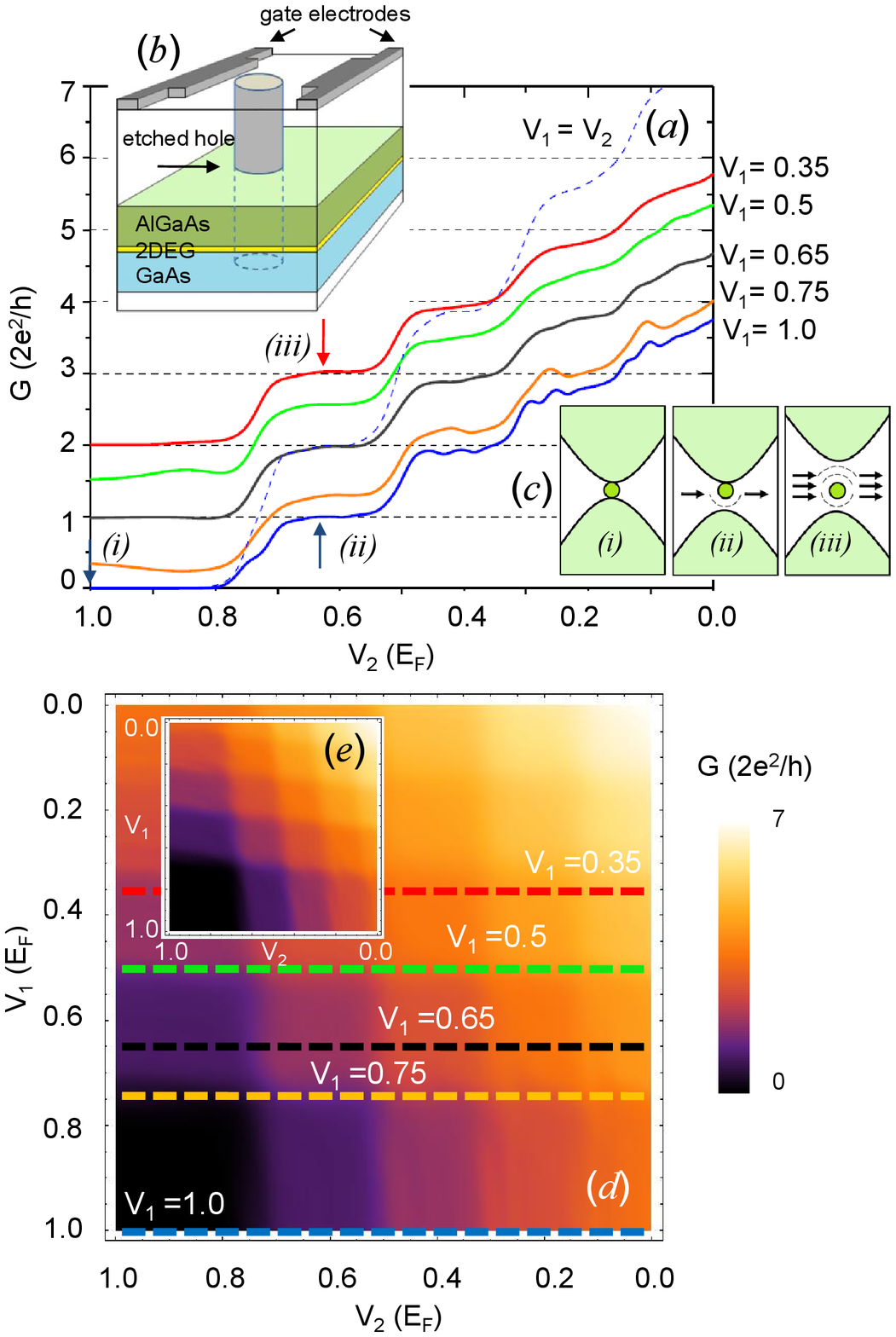}
\vspace{-0.3in}
\caption{\footnotesize (Color online) ({\it a}) Calculated conductance $G$ versus $V_2$ for different values of $V_1$ at zero magnetic field. Inset ({\it b}) shows a schematic diagram of the double QPC device.  As an illustration, inset ({\it c}) depicts three representative mode propagations in the device. In case ({\it i}) both QPCs are closed, case ({\it ii}) indicates that the propagation of only one mode through QPC$_1$ is allowed, and in case ({\it iii}), up to three modes are transmitted with both QPCs open.  ({\it d}) Intensity plot of $G$ obtained by sweeping $V_1$ and $V_2$ simultaneously.  
Inset ({\it e}) includes capacitive effect between the gates, see text for details.
} 
\end{figure}

\begin{figure}[!]
\vspace{-3.5cm}

\hspace*{3.0cm}\includegraphics[width=5.0in,height=7.0 in]{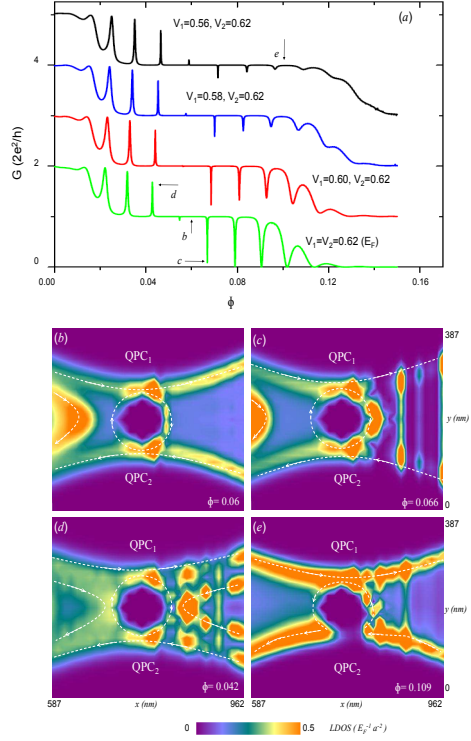}
\vspace{-0.8cm}

\caption{\footnotesize (Color online) ({\it a}) Conductance dependence on the magnetic flux $\Phi$ for different $V_1$ and $V_2$.  The plots are shifted for clarity. Two regimes are identified, a resonant behavior at low fields ($\Phi \lesssim 0.05$) and an anti-resonance at high fields ($\Phi> 0.05$). The period of the oscillations are presumably due to the Aharonov-Bohm effect. Figs. ({\it b})-({\it e}) are the plots of the local density of states (LDOS) at the central region of the double QPC device for different magnetic flux values and gate voltages. The white dashed lines indicate the different edge state paths.  }
\end{figure}

\end{document}